\documentclass[aps,twocolumn,superscriptaddress,groupedaddress]{revtex4}  % for review and submission

\usepackage{amssymb,amsmath,amsfonts,amsbsy}
\usepackage{color}
\usepackage{enumerate}
\usepackage{graphicx}
\usepackage{pifont}% http://ctan.org/pkg/pifont
\def\dA{{\rm d}A}

\def\dV{{\rm d}V}

\begin{document}

\title{Ensemble Average of Three-Dimensional Minkowski Tensors of a Gaussian Random Field in Redshift Space}

\author{Stephen Appleby}\email{stephen@kias.re.kr}
\affiliation{School of Physics, Korea Institute for Advanced Study, 85
Hoegiro, Dongdaemun-gu, Seoul, 02455, Korea}
\author{Joby P. K.}
\affiliation{School of Physical Sciences, National Institute of Science Education and Research, Jatni 752050, Odisha, India}
\author{Pravabati Chingangbam}
\affiliation{Indian Institute of Astrophysics, Koramangala II Block, Bangalore 560 034, India}
\author{Changbom Park}
\affiliation{School of Physics, Korea Institute for Advanced Study, 85
Hoegiro, Dongdaemun-gu, Seoul, 02455, Korea}

\begin{abstract}
We present the ensemble expectation values for the translation invariant, rank-2 Minkowski tensors in three-dimensions, for a linearly redshift space distorted Gaussian random field. The Minkowski tensors $W^{0,2}_{1}$, $W^{0,2}_{2}$ are sensitive to global anisotropic signals present within a field, and by extracting these statistics from the low redshift matter density one can place constraints on the redshift space distortion parameter $\beta = f/b$. We begin by reviewing the calculation of the ensemble expectation values $\langle W^{0,2}_{1} \rangle$, $\langle W^{0,2}_{2} \rangle $ for isotropic, Gaussian random fields, then consider how these results are modified by the presence of a linearly anisotropic signal. Under the assumption that all fields remain Gaussian, we calculate the anisotropic correction due to redshift space distortion in a coordinate system aligned with the line of sight, finding inequality between the diagonal elements of  $\langle W^{0,2}_{1} \rangle $, $\langle W^{0,2}_{2} \rangle $. The ratio of diagonal elements of these matrices provides a set of statistics that are sensitive only to the redshift space distortion parameter $\beta$. We estimate the Fisher information that can be extracted from the Minkowski tensors, and find $W^{0,2}_{1}$ is more sensitive to $\beta$ than $W^{0,2}_{2}$, and a measurement of $W^{0,2}_{1}$ accurate to $\sim 1\%$ can yield a $\sim 4\%$ constraint on $\beta$. Finally, we discuss the difference between using the matrix elements of the Minkowski tensors directly against measuring the eigenvalues. For the purposes of cosmological parameter estimation we advocate the use of the matrix elements, to avoid spurious anisotropic signals that can be generated by the eigenvalue decomposition. 
\end{abstract}

\maketitle

\section{Introduction} 

The scalar Minkowski functionals constitute a complete set of scalar valuations \citep{Hadwiger, nla.cat-vn1821482} that describe the morphology and topology of excursion sets of a field. They are used to quantify the global properties of excursion sets, such as volume, surface area, topology. Being scalar quantities, they are not sensitive to features such as orientation and alignment. 

The Minkowski tensors are a rank-p generalisation of the Minkowski functionals \citep{McMullen:1997,Alesker1999,2002LNP...600..238B,HugSchSch07,1367-2630-15-8-083028,JMI:JMI3331,Beisbart:2001gk}. In this work we focus on translation invariant rank-2 Minkowski tensors in three-dimensions. Specifically, we work with $W^{0,2}_{1}$ and $W^{0,2}_{2}$, which are integrals over the boundary of an excursion set, with integrands related to symmetric tensor products of the surface normal vector $\tilde{\bf{n}}$. Being directionally sensitive functions, their principle use is in the quantitative characterization of anisotropic patterns in a field. The mathematics underlying this class of descriptors has long been studied within the context of integral geometry \citep{nla.cat-vn2176896}, and they have found use in the material sciences \citep{PhysRevE.77.051805,Becker2003ComplexDS,Olszowka2006}.

The scalar Minkowski functionals have been extensively applied to two- and three-dimensional cosmological fields such as the temperature map of the Cosmic Microwave Background (CMB) and the density field as traced by galaxies in the low-redshift Universe \citep{1970Ap......6..320D,Adler,Hamilton:1986, Gott:1986uz,Ryden:1988rk,1989ApJ...340..625G,1989ApJ...345..618M,1992ApJ...387....1P,1991ApJ...378..457P, Matsubara:1994we,Matsubara:1995wj,Schmalzing:1995qn,Gott:2006yy,Gott:2008kk,0004-637X-529-2-795,2005ApJ...633....1P, Appleby:2017ahh, Appleby:2018jew, Chingangbam:2012wp}. By measuring the Minkowski functionals, we can reconstruct the composition, initial conditions and evolution of the Universe \citep{Choi:2010sx,Park:2009ja}. Specifically the amplitude and shape of the Minkowski Functional curves are sensitive to ratios of cumulants of the field, and hence provide measurements of the shape of the N-point functions. The mathematics underlying the Minkowski functionals of Gaussian random fields and their non-Gaussian generalisation has been presented in \citep{1970Ap......6..320D,Adler,Gott:1986uz,Hamilton:1986,Ryden:1988rk,Matsubara:1994wn,Matsubara:1994we,Matsubara:1995dv,Schmalzing97beyondgenus,1987ApJ...319....1G,1987ApJ...321....2W,1988ApJ...328...50M,Matsubara:1995ns,2000astro.ph..6269M,10.1111/j.1365-2966.2008.12944.x,Pogosyan:2009rg,Gay:2011wz,Codis:2013exa}, and the effect of redshift space distortion was elucidated in \citep{Matsubara:1995wj,Codis:2013exa}.

The scalar Minkowki functionals have been generalized to vector- and tensor-valued quantities on Euclidean space, and they are collectively known as Minkowski tensors \cite{McMullen:1997, Alesker1999, HugSchSch07}.  Rank-1 Minkowski tensors have been applied to study the sub-structure of galaxy clusters \citep{Beisbart:2001gk,Beisbart:2001vb}. Rank-2 Minkowski tensors can be sub-classified into translation covariant and translation invariant ones. Translation covariant Minkowski tensors have been used to study the sub-structure of spiral galaxies in \citep{2002LNP...600..238B}.  

More recently, the translation invariant rank-2 Minkowski tensors have been the subject of a series of recent cosmological applications by the authors. They were first applied to cosmology, particularly the CMB,  in \cite{Ganesan:2016jdk}. Further, their definition was generalized to  curved two-dimensional space, with emphasis on spaces of constant curvature such as the sphere, in \citep{Chingangbam:2017uqv}.  The ensemble expectation values for isotropic Gaussian random fields were derived in \citep{Chingangbam:2017uqv,Appleby:2018tzk} in two- and three-dimensions, respectively. The numerical reconstruction of these statistics are described in \citep{Ganesan:2016jdk,Appleby:2017uvb,Appleby:2018tzk}. They have been applied to cosmological data such as the CMB \citep{Ganesan:2016jdk,K.:2018wpn} and fields of the epoch of reionization \citep{Kapahtia:2017qrg,Kapahtia:2019ksk}.

In this work we extract the Minkowski tensors from the three-dimensional matter density field $\delta$. The matter distribution of the Universe is postulated to be statistically isotropic and homogeneous. For an isotropic field, the ensemble expectation value of the Minkowski tensors are proportional to the product of the identity matrix and scalar Minkowski functionals, and hence do not contain additional information. However, the Universe as revealed by observed galaxy catalogs is anisotropic due to the effect of redshift space distortion. 

The peculiar velocity of galaxies generates a spurious displacement in their apparent positions along the line-of-sight as inferred by redshift. On large scales, coherent infall of matter into overdensities gives rise to the Kaiser effect \citep{1987MNRAS.227....1K}, which deforms over/under-densities along the line of sight. On small scales, stochastic velocity components of objects bound to virialised structures generates the so-called Finger of God effect \citep{1972MNRAS.156P...1J}. These two extreme cases correspond to the linearized and fully non-linear limits of the impact of the peculiar velocity field. At intermediate, mildly non-linear scales one must treat both the velocity and density field as perturbatively non-Gaussian. 

The observed density field, modified by redshift space distortion, contains additional cosmological information compared to the real space (unobservable) counterpart, as the redshift space distortion is sourced by the velocity field. On large scales, by extracting information from the velocity field one can constrain the parameter $\beta = f /b$, where $f= \Omega_{\rm m}^{\gamma}$ is the growth factor and $b$ is the bias in the clustering amplitude of the density tracers. Methods of extracting this information include galaxy clustering \citep{Alam:2016hwk} and measurements of multipoles in the void-galaxy cross-correlation function \citep{Hamaus:2017dwj} . 

The redshift space distortion creates a coherent anisotropy parallel to the line of sight. As the Minkowski tensors are particularly sensitive to anisotropic signals, by extracting these statistics from the low redshift matter field, one can potentially obtain stringent constraints on the growth history of the density perturbations. In \citep{Appleby:2018tzk}, the authors measured the Minkowski tensors for a linearly redshift space distorted field as a precursor to extracting them from galaxy catalogs. However, to use these quantities for cosmological parameter estimation we first require a prediction for the ensemble expectation value of Gaussian random fields in redshift space.

In this work, we provide this prediction for three-dimensional Minkowski tensors extracted from a linearly redshift space distorted Gaussian matter density field. We show how these statistics are sensitive to the parameter $\beta$, and numerically confirm our analytic results. We focus on two particular Minkowski tensors - $W^{0,2}_{1}$, $W^{0,2}_{2}$ - that possess translational invariance, and calculate their ensemble average for an anisotropic spacetime. In section \ref{sec:2} we briefly review their ensemble average for isotropic fields. Then in section \ref{sec:3} we generalise our analysis to include a linear redshift space distortion correction, and numerically confirm our analytic results in \ref{sec:4}. We conclude in section \ref{sec:5}.

\section{Minkowski Tensors of Isotropic Gaussian Fields in Real Space} 
\label{sec:2}

We begin by reviewing the definition of the Minkowski tensors in three dimensions, and their ensemble expectation values for an isotropic, Gaussian random field. Although the ensemble expectation value for these statistics was briefly derived in \citep{Appleby:2018tzk} following the method used in \citep{Schmalzing:1997uc}, we perform the calculation explicitly in this section, as it will clarify the redshift space analysis in section \ref{sec:3}. Our calculation will roughly follow the scalar Minkowski functionals in redshift space \citep{Matsubara:1995wj,Codis:2013exa}, but we stress that the definitions and variables used in this work differ from those found in \citep{Codis:2013exa}. Analytic formulae for Minkowski tensors in two dimensions for Gaussian anisotropic fields will be discussed in \citep{priya:2019}.

The Minkowski tensors that we study are explicitly defined as  

\begin{eqnarray} \label{eq:w021} & & W^{0,2}_{1} \equiv {1 \over 6 V} \int_{\partial Q} {\bf \hat{n}}^{2} dA , \\ 
 \label{eq:w022} & & W^{0,2}_{2} \equiv {1 \over 3\pi V} \int_{\partial Q} G_{2} {\bf \hat{n}}^{2} dA , \end{eqnarray} 

\noindent where $Q$ is the excursion set defined by the volume enclosed by the boundary $\partial Q$. We focus on the density field $\delta$, and the boundary $\partial Q$ corresponds to the iso-density surface $\delta = \delta_{c}$ for some constant threshold $\delta_{c}$. $V$ is the total volume over which the field is defined, $dA$ is the infinitesimal area element on $\partial Q$, ${\bf \hat{n}}$ is the unit normal to the surface $\partial Q$ and $G_{\rm 2}$ is the mean curvature of the surface. ${\bf \hat{n}}^{2} = {\bf \hat{n}}\otimes {\bf \hat{n}} = (n_{i}n_{j} + n_{j}n_{i})/2$ is the symmetric tensor product. The statistics $W^{0,2}_{1}$, $W^{0,2}_{2}$ are $3\times 3$ matrices, and both the structure of the matrix and the magnitude of its components inform us about the properties of the field. They are translation, but {\it not} rotation, invariant and hence our results will depend on the coordinate system adopted.

If the field $\delta$ is drawn from a Gaussian ensemble, then we can predict the ensemble expectation value of $W^{0,2}_{1}$ and $W^{0,2}_{2}$ by integrating these quantities over a multi-variate Gaussian probability distribution $P(\vec{D})$, where $\vec{D}$ is a ten-dimensional vector composed of $(\delta, \nabla_{i} \delta, \nabla^{2}_{i}\delta, \nabla_{i}\nabla_{j}\delta)$ for $i,j=1,2,3$ and $j > i$. Throughout this work, double indices are not summed, and we use notation such that $i\neq j$.

We make the following field re-definitions, to create a set of dimensionless, unit variance variables --

\begin{eqnarray}
& &   x \equiv {\delta \over \sigma_{0}} ,  \qquad x_{i} \equiv {\sqrt{3}\nabla_{i} \delta \over \sigma_{1}},  \\
& &   x_{ii} = {\sqrt{5} \nabla^{2}_{i} \delta \over \sigma_{2}}, \qquad 
  x_{ij} = {\sqrt{15} \nabla_{i}\nabla_{j} \delta \over \sigma_{2}} , \end{eqnarray} 

\noindent where the cumulants are defined as 

\begin{equation} \sigma_{i}^{2} = {1 \over 2\pi^{2}} \int dk k^{2+2i} P_{\rm m} (k) , \end{equation} 

\noindent and $P_{\rm m}(k)$ is the power spectrum from which $\delta$ is drawn. We also define the normalised threshold $\nu = \delta_{\rm c}/\sigma_{0}$ and $X = \sqrt{x_{1}^{2} + x_{2}^{2} + x_{3}^{2}}$. 

The variables $x, x_{i}, x_{ii}, x_{ij}$ have the following correlations 

\begin{eqnarray}\nonumber  & &  \langle x^{2} \rangle = 1 , \qquad \langle x_{i}^{2} \rangle = 1 , \qquad \langle x^{2}_{ii} \rangle = 1 , \\ 
\nonumber & & \langle x^{2}_{ij} \rangle = 1 ,
\qquad  \langle x x_{ii} \rangle = -\gamma , \qquad  \langle x_{ii} x_{jj} \rangle = {1 \over 3} ,
 \end{eqnarray} 

\noindent where $\gamma = \sqrt{5}\sigma_{1}^{2}/(3\sigma_{0}\sigma_{2})$. The six variables $x_{i}, x_{ij}$ are Gaussian and have zero cross-correlations. The remaining four variables $x, x_{ii}$ are all correlated. 

In what follows we make use of the integral transformation 

\begin{eqnarray}
& & \nonumber   \int_{\partial Q} \dA\,F(x,x_{i},x_{ii},x_{ij}) = \\
& & \label{eq:it} \qquad \qquad  {\sigma_{1} \over \sigma_{0}} \int \dV \, X \, \delta_{\rm D}(x-\nu)\, F(x,x_{i},x_{ii},x_{ij}),
\end{eqnarray}

\noindent where $\delta_{D}(x)$ is the Dirac delta function and $F$ is an arbitrary function of $x,x_{i}, x_{ii}, x_{ij}$. Using equation ($\ref{eq:it}$), we can write $W_1^{0,2}$ and  $W_2^{0,2}$ as 

\begin{eqnarray}
   W_1^{0,2}  &=& {\sigma_{1}  \over 6\sqrt{3} \sigma_{0} V}\,\int  \dV  \ \delta_{\rm D}(x-\nu)  {{\mathcal M} \over X},\\
    W_2^{0,2} &=& {\sigma_{1} \over 3\sqrt{3} \pi \sigma_{0}V}\,\int  \dV \ \delta_{\rm D}(x-\nu)\ \frac{G_2 {\mathcal M}}{X}  ,
\end{eqnarray}

where the matrix $\mathcal M$ is given by
\begin{equation}
  \mathcal M=  \left(
  \begin{array}{ccc} 
    x_1^2 &  x_1 \,x_2 &  x_1 \,x_3 \\
    x_1 \,x_2 & x_2^2 &  x_2 \,x_3 \\
        x_1 \,x_3 & x_2 x_3  & x_3^2
  \end{array}\right).
  \label{eqn:M}
  \end{equation}

%%%%%%%%%%%%%%%%%%%%%%%%%%%%%%%%%%%%%%%%%%%%%%%%%%%%%%%%%%%%%%%%%%%%%%%%%
\subsection{Ensemble expectation values for isotropic Gaussian fields}
\label{sec:gaussian}

If the field is Gaussian, then we can write the ensemble expectation values of $W^{0,2}_{1}$ and $W^{0,2}_{2}$ as 

\begin{widetext} 

\begin{eqnarray}
 \label{eq:eev1} \langle  W_1^{0,2} \rangle   &=& {\sigma_{1} \over 6\sqrt{3} \sigma_{0} V}\, \int P(x,x_{i},x_{ii},x_{ij}) dx dx_{i} dx_{ii} dx_{ij} \int  \dV  \ \delta_{\rm D}(x-\nu)\ \frac{{\mathcal M}}{X} ,\\
\label{eq:eev2}     \langle W_2^{0,2} \rangle  &=& {\sigma_{1} \over 3\sqrt{3}\pi \sigma_{0}V}\, \int P(x,x_{i},x_{ii},x_{ij}) dx dx_{i} dx_{ii} dx_{ij} \int  \dV \ \delta_{\rm D}(x-\nu)\ \frac{G_2 {\mathcal M}}{X} ,
\end{eqnarray}

\end{widetext} 

\noindent where $P(x,x_{i},x_{ii},x_{ij})$ is the probability distribution function of the ten variables $x, x_{i}, x_{ii}, x_{ij}$ and the mean curvature $G_{2}$ will be defined shortly. Because we are focusing on statistically homogeneous fields, the integral $\int dV$ commutes with the other integrals, and cancels the volume factor in the denominators. 

For the Minkowski Tensor $W^{0,2}_{1}$, the integrand  in ($\ref{eq:eev1}$) contains no explicit $x_{ii}, x_{ij}$ dependence, so these variables can be integrated out and we can write 

\begin{equation}  \langle  W_{1}^{0,2}|_{lm} \rangle   = {\sigma_{1} \over 6\sqrt{3} \sigma_{0}}\, \int P(x,x_{i}) dx dx_{i}   \delta_{\rm D}(x-\nu) \frac{{\mathcal M_{lm}}}{X} , \end{equation} 

\noindent where 

\begin{equation}\label{eq:prob} P(x,x_{i})dx dx_{i} = {1 \over (2\pi)^{2}} e^{-{1 \over 2} (x^{2} + x_{1}^{2} + x_{2}^{2} + x_{3}^{2})} dx dx_{1}dx_{2}dx_{3} . \end{equation}

\noindent The integral ranges are $[-\infty, \infty]$ for $x,x_{1,2,3}$ and we have written the Minkowski tensor in component notation $W^{0,2}_{1}|_{lm}$ where $l,m = 1,2,3$. 

The $l \neq m$ components are zero, as the integrand ${\mathcal M}_{lm}/X$ is an odd function of $x_{l}$,$x_{m}$. The diagonal components read 

\begin{eqnarray}\nonumber & &  \langle  W_1^{0,2}|_{ll} \rangle   = {\sigma_{1} e^{-\nu^{2}/2} \over 6\sqrt{3} \sigma_{0}} {1 \over 4\pi^{2}} \, \int_{-\infty}^{\infty} dx_{1}  \int_{-\infty}^{\infty} dx_{2} \int_{-\infty}^{\infty} dx_{3} \times \\
& & \quad {x_{l}^{2} \over \sqrt{x_{1}^{2} + x_{2}^{2} + x_{3}^{2}}}e^{-{1 \over 2}(x_{1}^{2} + x_{2}^{2} + x_{3}^{2})} = {\sigma_{1} \over 9\sqrt{3} \pi \sigma_{0}}e^{-\nu^{2}/2} . \end{eqnarray}

\noindent The isotropy of the field has manifested as $\langle  W_1^{0,2} \rangle$ being proportional to the identity matrix -- all diagonal elements are equal.

The integrand of $\langle W^{0,2}_{2} \rangle$ contains all ten variables. To proceed, we de-correlate $x_{ii}$ and $x$ using the linear transformation 

\begin{equation} y_{ii} = {x_{ii} + \gamma x \over \sqrt{1-\gamma^{2}}} . \end{equation} 

\noindent The mean curvature $G_{2}$ is then given in terms of the field and its derivatives as 

\begin{widetext} 

\begin{eqnarray}
 \nonumber    G_2 &=&  \frac{\sqrt{3} \sigma_{2}}{2\sqrt{5}\sigma_{1}\left(x_1^2+x_2^2+x_3^2\right)^{3/2}}\bigg[ {2 \over \sqrt{3}}\bigg(x_1 x_2 x_{12} +
   x_2 x_3 x_{23}+ x_1 x_3 x_{13}\bigg) - \bigg(x_1^2(x_{22}+x_{33}) + x_2^2(x_{11}+x_{33}) + x_3^2(x_{11}+x_{22})\bigg) \bigg] \\ 
  \nonumber  &=&   \frac{\sqrt{3}\sigma_{2}}{2\sqrt{5}\sigma_{1}\left(x_1^2+x_2^2+x_3^2\right)^{3/2}}\bigg[ {2 \over \sqrt{3}}\bigg(x_1 x_2 x_{12} +
   x_2 x_3 x_{23}+ x_1 x_3 x_{13}\bigg) + 2 \gamma x \left( x_{1}^{2} + x_{2}^{2} + x_{3}^{2}\right) - \\
    \label{eq:G2} & & \qquad  \sqrt{1-\gamma^{2}}\bigg(x_1^2(y_{22}+y_{33}) + x_2^2(y_{11}+y_{33}) + x_3^2(y_{11}+y_{22})\bigg) \bigg] .
\end{eqnarray}

\end{widetext}

\noindent We can use the fact that all terms in the integrand of ($\ref{eq:eev2}$) that are odd in $x_{1},x_{2},x_{3}, x_{12},x_{13},x_{23}$ integrate to zero. We also show in appendix A that all terms proportional to $y_{11}, y_{22}, y_{33}$ in $G_{2}$ integrate to zero. The off-diagonal elements $l \neq m$ are zero. The only surviving non-zero contributions are those proportional to $x$ in ($\ref{eq:G2}$) -- 

\begin{widetext} 

\begin{eqnarray}
 \nonumber   \langle W^{0,2}_{2}|_{ll} \rangle &=& {\sigma_{2}\gamma  \over 12\sqrt{5}\pi^{3} \sigma_{0}} \int_{-\infty}^{\infty} dx \int_{-\infty}^{\infty} dx_{1} \int_{-\infty}^{\infty} dx_{2} \int_{-\infty}^{\infty} dx_{3} \delta_{\rm D}(x-\nu) x   e^{-{1 \over 2}\left(x^{2} + x_{1}^{2} +x_{2}^{2} + x_{3}^{2}\right)} {{\mathcal M|_{ll}} \over X^{2}} , \\
 &=&  { \sigma_{1}^{2}  \over 27 \pi \sqrt{2\pi}  \sigma^{2}_{0}} \nu e^{-\nu^{2}/2} .
\end{eqnarray}

\end{widetext} 

\noindent The Minkowski Tensor is again proportional to the identity matrix.

\section{Minkowski Tensors in Redshift Space} 
\label{sec:3}

We now repeat our analysis in redshift space. The statistics of the matter density field are modified by a distortion due to the peculiar velocity parallel to the line of sight. A redshift coordinate $\bf{s}$ is defined as 

\begin{equation} { \bf s} = {\bf x} + {{\bf \hat{x}} . {\bf v} \over H} , \end{equation} 

\noindent where $\bf{x}$ is the real space coordinate $\bf{\hat{x}}$ is the unit vector parallel to the line of sight, $\bf{v}$ is the velocity field and $H$ is the Hubble parameter. The corresponding density field in redshift space $\delta^{\rm (s)}$ is related to its real-space counterpart $\delta^{(\rm x)}$ according to 

\begin{equation} \label{eq:n1} \delta^{(\rm s)}({\bf k}) = (1 + f \mu^{2}) \delta^{(\rm x)} ({\bf k}) , \end{equation} 

\noindent where $f = d\ln D/d\ln a \simeq \Omega_{\rm m}^{\gamma}$ is the linear growth factor and $\mu = {\bf k}. {\bf 
\hat{x}}/|k|$ is the cosine of the angle between the line of sight and wavenumber ${\bf k}$. The presence of the angular dependence $\mu$ in ($\ref{eq:n1}$) breaks the isotropy of the field, and modifies the field cumulants both parallel and perpendicular to the line of sight. If we are measuring the matter field via a biased tracer, the $f$ term in ($\ref{eq:n1}$) is replaced by $\beta = f/b$, where $b$ is the (linear) bias factor between the observational tracer and the underlying matter field. Our goal in this section is to estimate the ensemble expectation value of the Minkowski tensors $W^{0,2}_{1}$, $W^{0,2}_{2}$ that can be extracted from the redshift space distorted field $\delta^{(\rm s)}$. 

We calculate $\langle W^{0,2}_{1} \rangle$, $\langle W^{0,2}_{2} \rangle$ following the previous calculation as closely as possible. In what follows, we take the line of sight to be parallel to the $x_{3}$ direction, breaking the symmetry along this axis.  We define the following unit variance variables 

\begin{eqnarray} & & x \equiv {\delta \over \sigma}, \qquad x_{I} = {\sqrt{3} \nabla_{I} \delta \over \sigma_{1\perp}}, \qquad x_{3} = {\sqrt{3} \nabla_{3} \delta \over \sigma_{1\parallel}}, \\
& & x_{II} = {\sqrt{5} \nabla_{I}^{2} \delta \over \sigma_{2\perp}}, \qquad x_{33} = {\sqrt{5}\nabla_{3}^{2} \delta \over \sigma_{2 \parallel}}, \\
& &  x_{IJ} = {\sqrt{15}\nabla_{I}\nabla_{J} \delta \over \sigma_{2\perp}}, \qquad x_{I3} = {\sqrt{15}\nabla_{I}\nabla_{3} \delta \over \sigma_{2\times}},  \end{eqnarray}  

\noindent where $I, J =1,2$ and $I \neq J$, $J > I$. We have defined 

\begin{eqnarray} \label{eq:sigs}  & & \sigma^{2} = A_{0}\sigma_{0}^{2} , \qquad 
 \sigma_{1 \perp}^{2} = A_{1\perp} \sigma_{1}^{2} , \qquad  \sigma_{1\parallel}^{2} = A_{1\parallel} \sigma_{1}^{2} , \\
 \nonumber  & & \sigma_{2\perp}^{2} = A_{2\perp} \sigma_{2}^{2} , \qquad  \sigma_{2\parallel}^{2} = A_{2\parallel}  \sigma_{2}^{2}  , \qquad \sigma_{2\times}^{2} = A_{2\times} \sigma_{2}^{2} , \end{eqnarray} 
 
 \noindent where the $A$ factors are the modifications to the cumulants due to the effect of redshift space distortion, normalised such that all $\sigma$ terms in ($\ref{eq:sigs}$) approach their isotropic limits as $f/b \to 0$. Explicitly the $A$ factors are given by 
 
 \begin{eqnarray} 
  \nonumber & & A_{0} = 1 + {2 \beta \over 3} + {\beta^{2} \over 5}, \quad A_{1\perp} = 1 + {6 \beta \over 15} + {3\beta^{2} \over 35} , \\
\nonumber & & A_{1\parallel} = 1 + {6 \beta \over 5} + {3 \beta^{2} \over 7}, \quad A_{2\parallel} = 1 + {10 \beta \over 7} + {5\beta^{2} \over 9} , \\ 
\nonumber  & & A_{2\perp} = 1 + {2 \beta \over 7} + {\beta^{2} \over 21}, \quad
  A_{2\times} = 1 + {6\beta \over 7} + {5 \beta^{2} \over 21} ,
\end{eqnarray} 

\noindent All $A$ terms approach unity as $\beta \to 0$. 

The six variables $x_{I}, x_{3}, x_{IJ}, x_{I3}$ are uncorrelated. The remaining terms possess the following correlations 

\begin{eqnarray} & &  \langle x x_{II} \rangle = -\gamma_{\perp} , \qquad \langle x x_{33} \rangle = -\gamma_{\parallel} , \\
& & \langle x_{II}x_{JJ} \rangle = {1 \over 3} , \qquad \langle x_{II}x_{33} \rangle = {\sigma_{2\times}^{2} \over 3\sigma_{2\parallel}\sigma_{2\perp}}  , \end{eqnarray} 

\noindent where $\gamma_{\parallel} = \sqrt{5}\sigma_{1\parallel}^{2}/(3\sigma \sigma_{2\parallel})$, $\gamma_{\perp} = \sqrt{5}\sigma_{1\perp}^{2}/(3\sigma \sigma_{2\perp})$.  We can repeat our real space analysis for $W^{0,2}_{1}$, accounting for the anisotropy in the $x_{3}$ direction. The probability distribution $P(x,x_{1},x_{2},x_{3})$ is unchanged from the isotropic case -- equation ($\ref{eq:prob}$) -- and only the integrand changes. Specifically, the matrix $\tilde{\mathcal{M}}$ is now given by

\begin{equation}
  \tilde{\mathcal{M}} =  \left(
  \begin{array}{ccc} 
    x_1^2 &  x_1 \,x_2 &  \lambda x_1 \,x_3 \\
    x_1 \,x_2 & x_2^2 &  \lambda x_2 \,x_3 \\
        \lambda x_1 \,x_3 & \lambda  x_2 x_3  &\lambda^{2} x_3^2
  \end{array}\right) ,
  \label{eqn:Mrsd}
  \end{equation}

\noindent where $\lambda^{2} = \sigma_{1\parallel}^{2}/\sigma_{1\perp}^{2}$ and 

\begin{equation}  \langle  W_{1}^{0,2}|_{lm} \rangle   = {\sigma_{1 \perp} \over 6\sqrt{3} \sigma} \, \int P(x,x_{i}) dx dx_{i}   \delta_{\rm D}(x-\nu) {\tilde{\mathcal{M}}_{lm} \over \tilde{X}} , \end{equation} 

\noindent where $\tilde{X} = \sqrt{x_{1}^{2} + x_{2}^{2} + \lambda^{2} x_{3}^{2}}$. 

The same argument used for the isotropic field can be used to set the off-diagonal elements to zero -- the integrand $\tilde{\mathcal M}_{lm}/\tilde{X}$ is an odd function of $x_{1},x_{2},x_{3}$ for $l \neq m$. What remains is the diagonal elements. The integrals for the $11$, $22$ components do not individually yield an analytic result, but we can integrate the $11 + 22$ component and use the symmetry in this two dimensional space to fix the individual elements. We find 

\begin{widetext}

\begin{eqnarray}\label{eq:m1} & &  \langle W^{0,2}_{1}|_{II} \rangle =  {\sigma_{1\perp} \over 24 \sqrt{3} \pi \sigma}\left[ {(2\lambda^{2}-1)\cosh^{-1}\left(2\lambda^{2}-1\right) \over (\lambda^{2}-1)^{3/2}} - {2\lambda \over \lambda^{2}-1}  \right] e^{-\nu^{2}/2} , \\
\label{eq:m2} & & \langle W^{0,2}_{1}|_{33} \rangle = {\sigma_{1\perp} \over 6 \sqrt{3} \pi \sigma}\left({\lambda^{2} \over \lambda^{2}-1}\right) \left( \lambda - {\cosh^{-1} \lambda \over \sqrt{\lambda^{2}-1}}\right) e^{-\nu^{2}/2} , \\
\label{eq:m3} & & \langle W^{0,2}_{1}|_{IJ} \rangle = \langle W^{0,2}_{1}|_{I3} \rangle = 0 ,
\end{eqnarray} 

\end{widetext}

\noindent valid for $\lambda > 1$. The isotropic limit corresponds to $\beta \to 0$, in which case $\lambda \to 1$, $\sigma \to \sigma_{0}$, $\sigma_{1\perp}, \sigma_{1\parallel} \to \sigma_{1}$, and 

\begin{equation}   \lim_{\beta \to 0} \langle W^{0,2}_{1}|_{II} \rangle = \lim_{\beta \to 0} \langle W^{0,2}_{1}|_{33} \rangle = {\sigma_{1} \over 9\sqrt{3} \pi \sigma_{0}} e^{-\nu^{2}/2}  ,
\end{equation}

\noindent in agreement with the calculation for isotropic fields in section \ref{sec:gaussian}. 

Next we calculate the ensemble expectation value of $W^{0,2}_{2}$ for a linearly redshift space distorted field. The mean curvature $G_{2}$ is now given by

 \begin{widetext} 

\begin{eqnarray}
 \nonumber    G_2 &=&  {\sqrt{3}  \over 2\sqrt{5}\sigma_{1 \perp}\left(x_1^2+x_2^2+\lambda^{2} x_3^2\right)^{3/2}}\left[  {2 \over \sqrt{3}} \left( \sigma_{2\perp} x_1 x_2 x_{12} +
   \lambda \sigma_{2\times} x_2 x_3 x_{23} + \lambda \sigma_{2\times } x_1 x_3 x_{13} \right) \right. \\
    & & \qquad \left. - \left( x_1^2( \sigma_{2\perp} x_{22}+ \sigma_{2\parallel} x_{33}) + x_2^2(\sigma_{2\perp} x_{11}+ \sigma_{2\parallel} x_{33}) + \lambda^{2}\sigma_{2\perp} x_3^2(x_{11}+x_{22}) \right)  \right]  .
\end{eqnarray}

\end{widetext}

\noindent As before, we de-correlate $x_{II}$, $x_{33}$ and $x$ via the linear transformations 

\begin{eqnarray}  y_{II} &=& {x_{II} + \gamma_{\perp} x \over \sqrt{1-\gamma^{2}_{\perp}}} , \\
 y_{33} &=& {x_{33} + \gamma_{\parallel} x \over \sqrt{1-\gamma_{\parallel}^{2}}} . \end{eqnarray}

Similarly to the isotropic case, all terms in $G_{2}$ involving $x_{IJ}, x_{I3}$ integrate to zero when integrated over the joint probability distribution $P(x,x_{I},x_{3},x_{IJ},x_{I3},y_{II},y_{33})$. We show in appendix A that the terms in $G_{2}$ proportional to $y_{II}$ and $y_{33}$ also integrate to zero. The surviving elements of $G_{2}$ are 

\begin{widetext} 

\begin{eqnarray}   \tilde{G}_{2} &=&    {\sqrt{3} x \over 2\sqrt{5}\sigma_{1 \perp}  \left(x_1^2+x_2^2+\lambda^{2} x_3^2\right)^{3/2}} \left[ \left( x_{1}^{2} + x_{2}^{2} \right) \left( \sigma_{2\perp}\gamma_{\perp} + \sigma_{2\parallel}\gamma_{\parallel}\right) + 2\lambda^{2} x_{3}^{2} \sigma_{2\perp}\gamma_{\perp} \right] ,   \\
 &=& { \sigma_{1\perp} x \over 2\sqrt{3}\sigma  \left(x_1^2+x_2^2+\lambda^{2} x_3^2\right)^{3/2}} \left[ \left( x_{1}^{2} + x_{2}^{2} \right) \left( 1 + \lambda^{2} \right) + 2\lambda^{2} x_{3}^{2}  \right] ,
\end{eqnarray} 

\noindent and the corresponding non-zero components of the Minkowski Tensor $W^{0,2}_{2}$ are 

\begin{eqnarray}
    \langle W^{0,2}_{2}|_{II} \rangle &=& {\sigma_{1\perp}^{2} \nu e^{-\nu^{2}/2} \over 72 \pi^{3} \sigma^{2}}   \int_{-\infty}^{\infty} dx_{1} \int_{-\infty}^{\infty} dx_{2} \int_{-\infty}^{\infty} dx_{3}   e^{-{1 \over 2}\left(x_{1}^{2} +x_{2}^{2} + x_{3}^{2}\right)} {\tilde{\mathcal{M}}|_{II} \over \tilde{X}^{4}} \left[ (x_{1}^{2} + x_{2}^{2})(1+\lambda^{2}) + 2\lambda^{2} x_{3}^{2}\right] , \\
    \langle W^{0,2}_{2}|_{33} \rangle &=& {\sigma_{1\perp}^{2} \nu e^{-\nu^{2}/2} \over 72 \pi^{3} \sigma^{2}}   \int_{-\infty}^{\infty} dx_{1} \int_{-\infty}^{\infty} dx_{2} \int_{-\infty}^{\infty} dx_{3}   e^{-{1 \over 2}\left(x_{1}^{2} +x_{2}^{2} + x_{3}^{2}\right)} {\tilde{\mathcal{M}}|_{33} \over \tilde{X}^{4}} \left[ (x_{1}^{2} + x_{2}^{2})(1+\lambda^{2}) + 2\lambda^{2} x_{3}^{2}\right] .
\end{eqnarray}

\noindent The integrals admit analytic solutions 

\begin{eqnarray}
 \label{eq:m4}   \langle W^{0,2}_{2}|_{II} \rangle &=& {\sigma_{1\perp}^{2} \over 72 \pi \sqrt{2\pi} \sigma^{2}} {\left[(\lambda^{2}-2)(\lambda^{2}-1)^{1/2} + \lambda^{4} \tan^{-1}\sqrt{\lambda^{2}-1}  \right] \over (\lambda^{2}-1)^{3/2} }\, \nu e^{-\nu^{2}/2} ,  \\
 \label{eq:m5}    \langle W^{0,2}_{2}|_{33} \rangle &=& {\sigma_{1\perp}^{2}  \over 36 \pi \sqrt{2\pi} \sigma^{2}}  {\lambda^{2} \left[ (\lambda^{2}-1)^{1/2} + (\lambda^{2}-2) \tan^{-1}\sqrt{\lambda^{2}-1} \right] \over (\lambda^{2}-1)^{3/2} } \, \nu e^{-\nu^{2}/2} , \\
 \label{eq:m6} \langle W^{0,2}_{2}|_{IJ} \rangle &=& \langle W^{0,2}_{2}|_{I3} \rangle = 0 .
\end{eqnarray}

\end{widetext}

\noindent In the isotropic limit, $\beta \to 0$, $\lambda \to 1$ and $\sigma \to \sigma_{0}$, $\sigma_{2\perp}, \sigma_{2\parallel}, \sigma_{2\times} \to \sigma_{2}$, $\sigma_{1\perp}, \sigma_{1\parallel} \to \sigma_{1}$ and 

\begin{equation} \lim_{\beta \to 0} \langle W^{0,2}_{2}|_{II} \rangle = \lim_{\beta \to 0} \langle W^{0,2}_{2}|_{33} \rangle = {\sigma_{1}^{2} \over 27 \pi \sqrt{2\pi} \sigma_{0}^{2}} \nu e^{-\nu^{2}/2} . \end{equation} 

\noindent For both $W^{0,2}_{1}$ and $W^{0,2}_{2}$, the effect of redshift space distortion is to generate inequality between the diagonal elements $II$ and $33$, where $33$ is the direction in which isotropy is broken.

To review our results, we write the expectation values of $W^{0,2}_{1}$, $W^{0,2}_{2}$ in redshift space as 

\begin{eqnarray} \langle W^{0,2}_{1}|_{ii}\rangle &=&  a_{1}|_{i} e^{-\nu^{2}/2} H_{0}(\nu) , \\
 \langle W^{0,2}_{2}|_{ii}\rangle &=&  a_{2}|_{i} e^{-\nu^{2}/2} H_{1}(\nu) , \\
\end{eqnarray}

\noindent where $H_{0,1}(\nu)$ are the Hermite polynomials $H_{0}(\nu) = 1$, $H_{1}(\nu) = \nu$ and 

\begin{widetext} 

\begin{eqnarray}\label{eq:co1} & & a_{1}|_{I} =  {\sigma_{1\perp} \over 24 \sqrt{3} \pi \sigma}\left[ {(2\lambda^{2}-1)\cosh^{-1}\left(2\lambda^{2}-1\right) \over (\lambda^{2}-1)^{3/2}} - {2\lambda \over \lambda^{2}-1}  \right] , \\
\label{eq:co2} & & a_{1}|_{3} = {\sigma_{1\perp} \over 6 \sqrt{3} \pi \sigma}\left({\lambda^{2} \over \lambda^{2}-1}\right) \left( \lambda - {\cosh^{-1} \lambda \over \sqrt{\lambda^{2}-1}}\right) , \\
\label{eq:co3} & & a_{2}|_{I} =  {\sigma_{1\perp}^{2} \over 72 \pi \sqrt{2\pi} \sigma^{2}} {\left[(\lambda^{2}-2)(\lambda^{2}-1)^{1/2} + \lambda^{4} \tan^{-1}\sqrt{\lambda^{2}-1}  \right] \over (\lambda^{2}-1)^{3/2} } , \\
\label{eq:co4} & & a_{2}|_{3} =  {\sigma_{1\perp}^{2}  \over 36 \pi \sqrt{2\pi} \sigma^{2}}  {\lambda^{2} \left[ (\lambda^{2}-1)^{1/2} + (\lambda^{2}-2) \tan^{-1}\sqrt{\lambda^{2}-1} \right] \over (\lambda^{2}-1)^{3/2} } ,
\end{eqnarray}

\noindent for $I=1,2$ and 

\begin{equation} \lambda^{2} = {\sigma_{1\parallel}^{2} \over \sigma_{1\perp}^{2}} = {35 + 42\beta   + 15 \beta^{2} \over 35 + 14\beta  + 3 \beta^{2}} . \end{equation} 

\end{widetext}

The theoretical expectation values ($\ref{eq:co1}-\ref{eq:co4}$) are the main results of this work. For an isotropic field the ensemble expectation value of the Minkowski tensors is proportional to the identity matrix. For a density field in which anisotropy is generated by linear redshift space distortion, the ensemble expectation values of $W^{0,2}_{1}$ and $W^{0,2}_{2}$ are diagonal but the matrix components are no longer equal. The results obtained here are valid in a coordinate system in which the line of sight is aligned with the $x_{3}$ axis; the Minkowski tensors are not coordinate invariant. 

From the diagonal elements ($\ref{eq:co1}-\ref{eq:co4}$) we can define a set of statistics that are sensitive only to the redshift space distortion signal. The ratios 

\begin{widetext} 

\begin{eqnarray} \label{eq:th0} & & \Theta_{1}|_{I} = {a_{1}|_{I} \over a_{1}|_{3}} = { (2\lambda^{2}-1)\cosh^{-1}(2\lambda^{2}-1) - 2\lambda \sqrt{\lambda^{2}-1} \over 4\lambda^{2} \left(\lambda \sqrt{\lambda^{2}-1} - \cosh^{-1}\lambda \right)} , \\
\label{eq:th1} & & \Theta_{2}|_{I} = {a_{2}|_{I} \over a_{2}|_{3}} = {\lambda^{4} \tan^{-1} \sqrt{\lambda^{2}-1} + (\lambda^{2}-2)\sqrt{\lambda^{2}-1}  \over 2\lambda^{2} \left[ (\lambda^{2}-2)\tan^{-1}\sqrt{\lambda^{2}-1} + \sqrt{\lambda^{2}-1} \right]}  ,  \end{eqnarray} 

\end{widetext} 

\noindent are sensitive only to $\beta = f/b$; all other cosmological parameter dependence drops out. Hence by extracting the Minkowski tensors from the matter density field (as traced by galaxies, for example) we can obtain a constraint on the redshift space distortion parameter $\beta$ by measuring $\Theta_{1|I}$ and $\Theta_{2|I}$.

\begin{table}
\begin{center}
%\captionof{table}{}\label{tab:1} 
 \begin{tabular}{||c | c ||}
 \hline
 Parameter & Fiducial Value \\ [0.5ex] 
 \hline\hline
 $\Omega_{\rm m}$ & $0.26$   \\ 
 \hline
 $w_{\rm de}$ & $-1$ \\
 \hline
 $n_{\rm s}$ & $0.96$   \\ 
 \hline
 $h$ & $0.72$   \\
 \hline
 $R_{\rm G}$ & $20 {\rm Mpc}$ \\ 
  \hline 
  $b$ & $1.0$ \\
  \hline
  $\beta$ & $0.48$ \\
  \hline
\end{tabular}
\caption{\label{tab:I}Fiducial parameters used to generate the Gaussian Random Fields. $R_{\rm G}$ is the Gaussian smoothing scale, and $\beta = \Omega_{\rm m}^{6/11}/b$. }
\end{center} 
\end{table}

\section{Numerical Analysis} 
\label{sec:4}

We confirm our analytic predictions by generating isotropic Gaussian random fields, as well as anisotropic fields that suffer from the linear redshift space distortion. For the isotropic case, we generate a Gaussian field in Fourier space using a $\Lambda$CDM linear matter power spectrum. We use cosmological parameters presented in table \ref{tab:I}, and a box of size $V = 1024^{3} {\rm Mpc}^{3}$ with resolution $\Delta = 2 {\rm Mpc}$.

For the anisotropic random fields, we first generate an isotropic field, then Fourier transform it, and apply the following transformation in Fourier space -- 

\begin{equation} \delta({\bf k}) \to \left( 1 + \beta {k_{z}^{2} \over k^{2}} \right) \delta({\bf k}) . \end{equation} 

\begin{figure*}
  \includegraphics[width=0.48\textwidth]{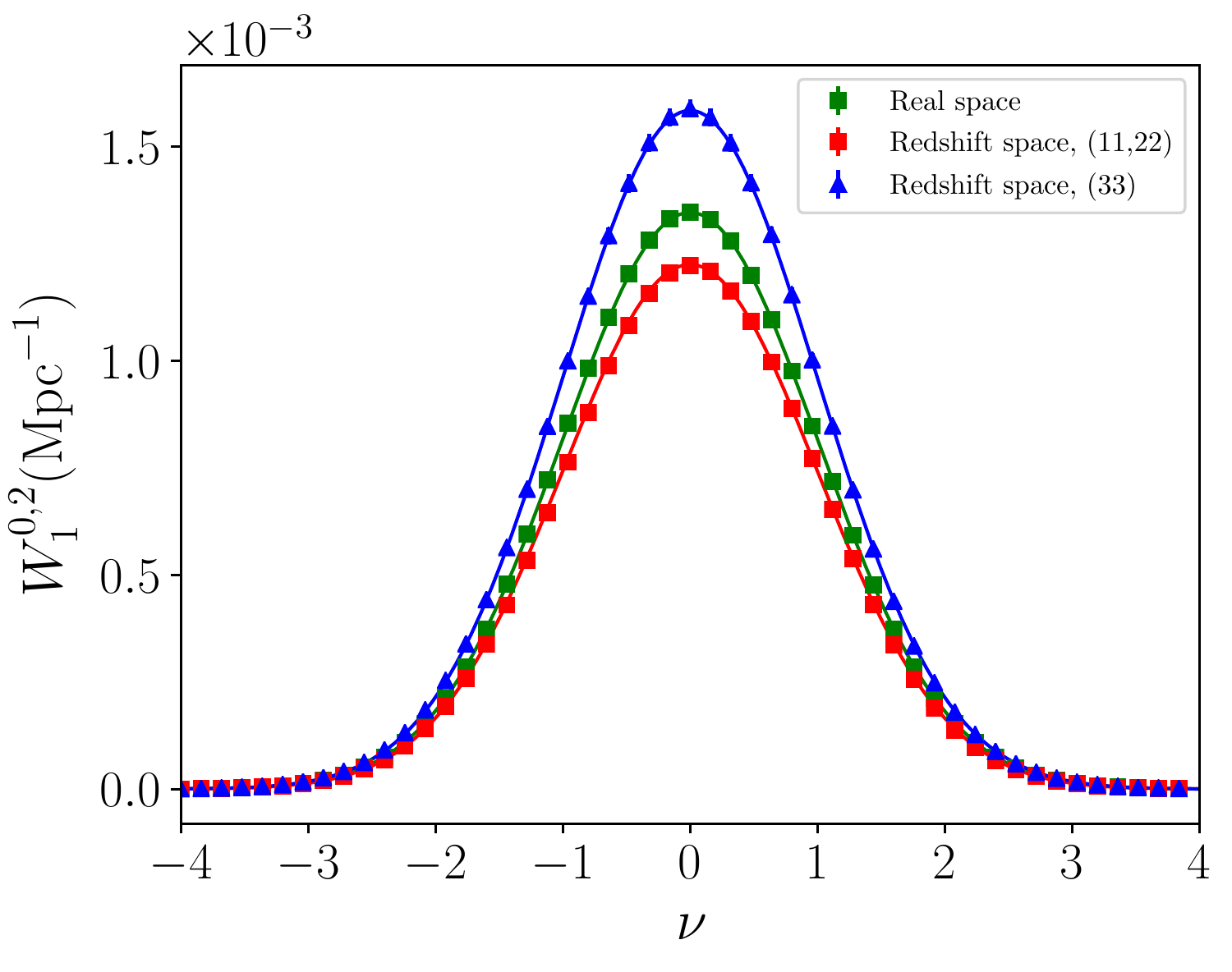}
      \includegraphics[width=0.48\textwidth]{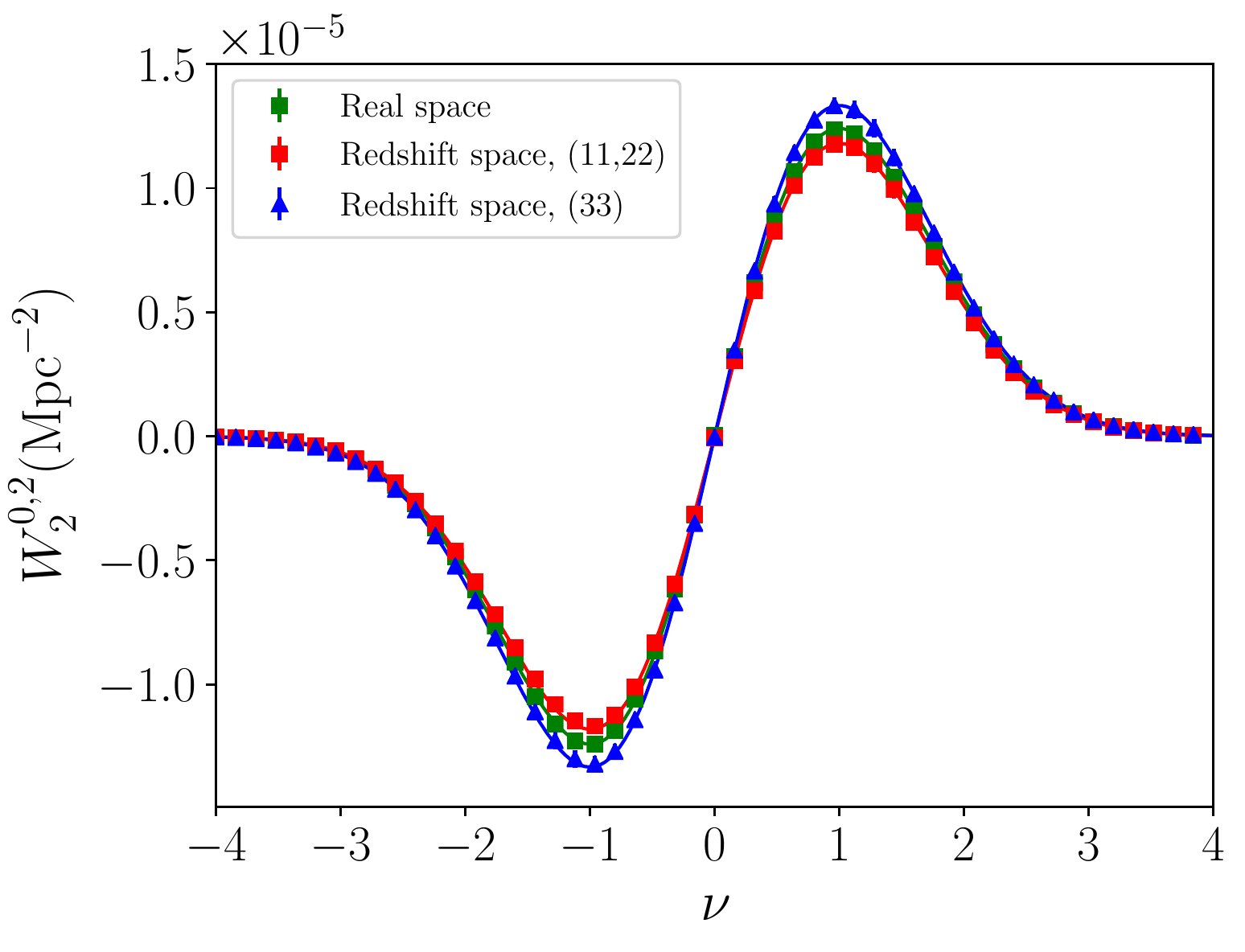}
  \includegraphics[width=0.48\textwidth]{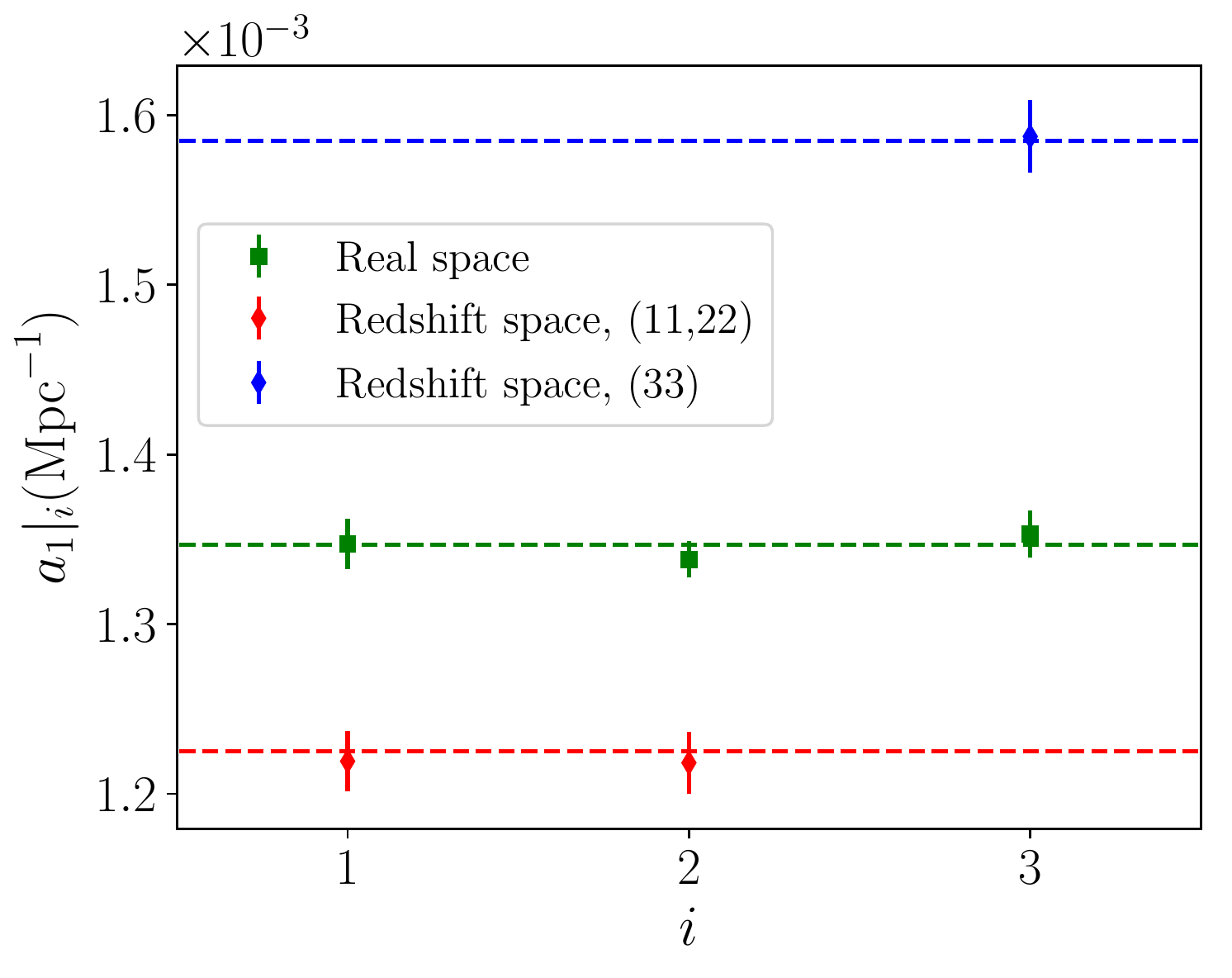}
  \includegraphics[width=0.48\textwidth]{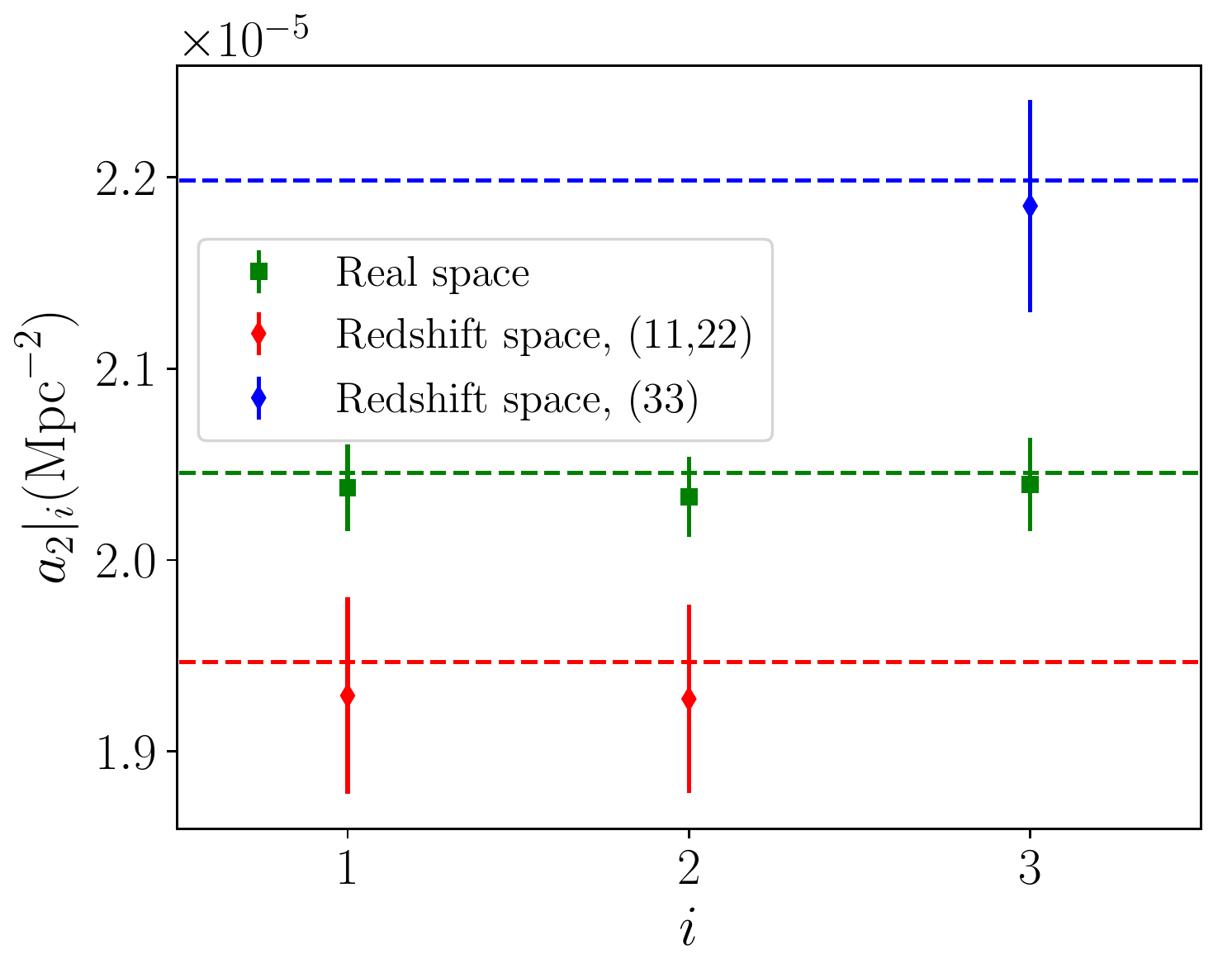}
  \caption{[Top panel] The diagonal components of the Minkowski tensors $W^{0,2}_{1}$ (left) and $W^{0,2}_{2}$ (right). The points/error bars are the mean and rms as measured from $N_{\rm real} = 100$ Gaussian realisations, and the solid lines are the predictions presented in the text. The green squares/lines correspond to isotropic fields, and the blue/red points/lines are the $33$ and $11, 22$ components of an anisotropic Gaussian field  [Bottom panel] The amplitude of the top panels, extracted via the relations ($\ref{eq:he1}$,$\ref{eq:he2}$). The color scheme is consistent with the top panels. In redshift space, the amplitudes are significantly modified due to the global anisotropy generated by the velocity field. }
  \label{fig:1}
\end{figure*}

\noindent We then apply a Gaussian smoothing kernel of width $R_{\rm G}$ and inverse Fourier transform. Using the methods outlined in \citep{Appleby:2018tzk}, we numerically reconstruct the Minkowski tensors $W^{0,2}_{1}$, $W^{0,2}_{2}$ from these fields. In the top panels of Figure \ref{fig:1} we present the mean and standard deviations of the diagonal components of $W^{0,2}_{1}$, $W^{0,2}_{2}$, extracted from $N_{\rm real} = 100$ realisations of isotropic and anisotropic random fields. We measure the statistics at $N_{\nu} = 50$ values equi-spaced over the range $\nu_{\rm min} < \nu < \nu_{\rm max}$ for $\nu_{\rm min} = -4$, $\nu_{\rm max} = 4$, these points are presented as green squares for the isotropic field and blue triangles/red squares for the anisotropic field. The blue/red solid lines in the panels show the analytic predictions ($\ref{eq:m2},\ref{eq:m1}$), ($\ref{eq:m5},\ref{eq:m4}$). The green line corresponds to the isotropic limit of the expressions ($\ref{eq:m1},\ref{eq:m2}$) and ($\ref{eq:m4},\ref{eq:m5}$). One can observe close agreement between our numerical reconstruction and the theoretical expectations, with the theoretical prediction reproduced to accuracy $< 0.5\%$ for $W^{0,2}_{1}$ and $< 1\%$ for $W^{0,2}_{2}$.

For each realisation, we measure the components of $W^{0,2}_{1}$, $W^{0,2}_{2}$ as a function of threshold $\nu$ and then extract the amplitudes $a_{1}|_{I}$, $a_{1}|_{3}$, $a_{2}|_{I}$, $a_{2}|_{3}$ by generating a smooth spline between the $N_{\nu} = 50$ measured points and numerically integrating 

\begin{eqnarray}\label{eq:he1} a_{1}|_{i} &=& {1 \over \sqrt{2\pi}} \int_{\nu_{\rm min}}^{\nu_{\rm max}} W^{0,2}_{1}|_{ii} d\nu  ,  \\
\label{eq:he2} a_{2}|_{i} &=&  {1 \over \sqrt{2\pi}} \int_{\nu_{\rm min}}^{\nu_{\rm max}} \nu W^{0,2}_{2}|_{ii}  d\nu , \end{eqnarray}

\noindent where we have used the orthogonality property of the Hermite polynomials 

\begin{equation} \int_{-\infty}^{\infty} H_{n}(\nu) H_{m}(\nu) e^{-\nu^{2}/2} d\nu = \sqrt{2\pi} n! \delta_{nm} . \end{equation}

\noindent Technically the integrals ($\ref{eq:he1},\ref{eq:he2}$) only reproduce the amplitudes $a_{1}|_{i}$, $a_{2}|_{i}$ when the integration range is $[-\infty, \infty]$, but we have checked that $[-4,4]$ is a sufficiently broad range for our purpose due to the exponential decrease in $W^{0,2}_{1}$, $W^{0,2}_{2}$ at large $|\nu|$. After measuring $a_{1}|_{i}$, $a_{2}|_{i}$ one can take the ratios $a_{1}|_{I}/a_{1}|_{3}$, $a_{2}|_{I}/a_{2}|_{3}$ to obtain $\Theta_{1}|_{I}$, $\Theta_{2}|_{I}$. In the bottom panels of Figure \ref{fig:1} we present the amplitudes numerically extracted from the data (points/error bars) and the theoretical expectation values ($\ref{eq:co1}-\ref{eq:co4}$) presented in this work (dashed lines). The agreement between prediction and numerical reconstructions is better than $< 1\%$ for all $a_{1}|_{i}$, $a_{2}|_{i}$ terms.  

\begin{figure}
  \includegraphics[width=0.48\textwidth]{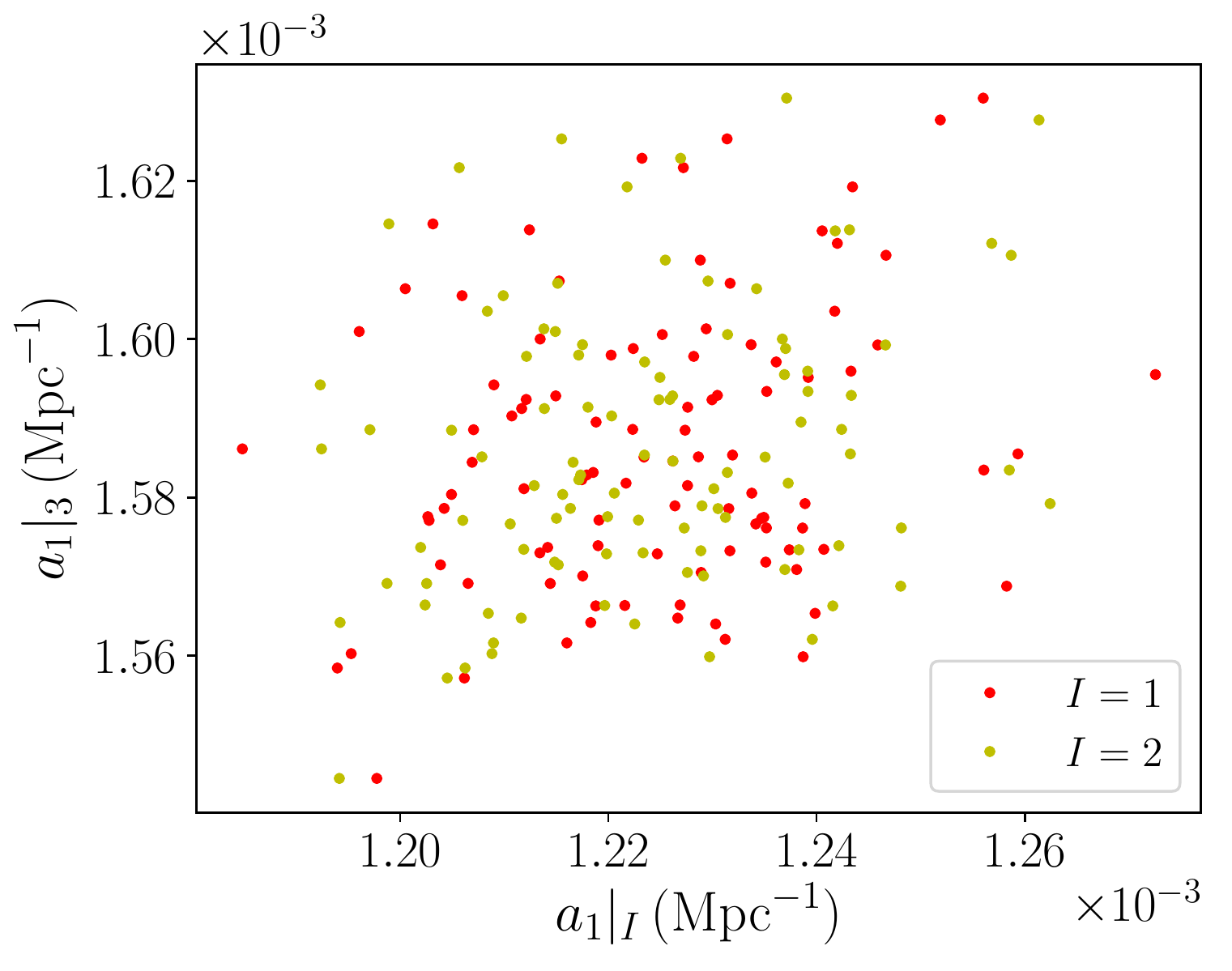}
      \includegraphics[width=0.48\textwidth]{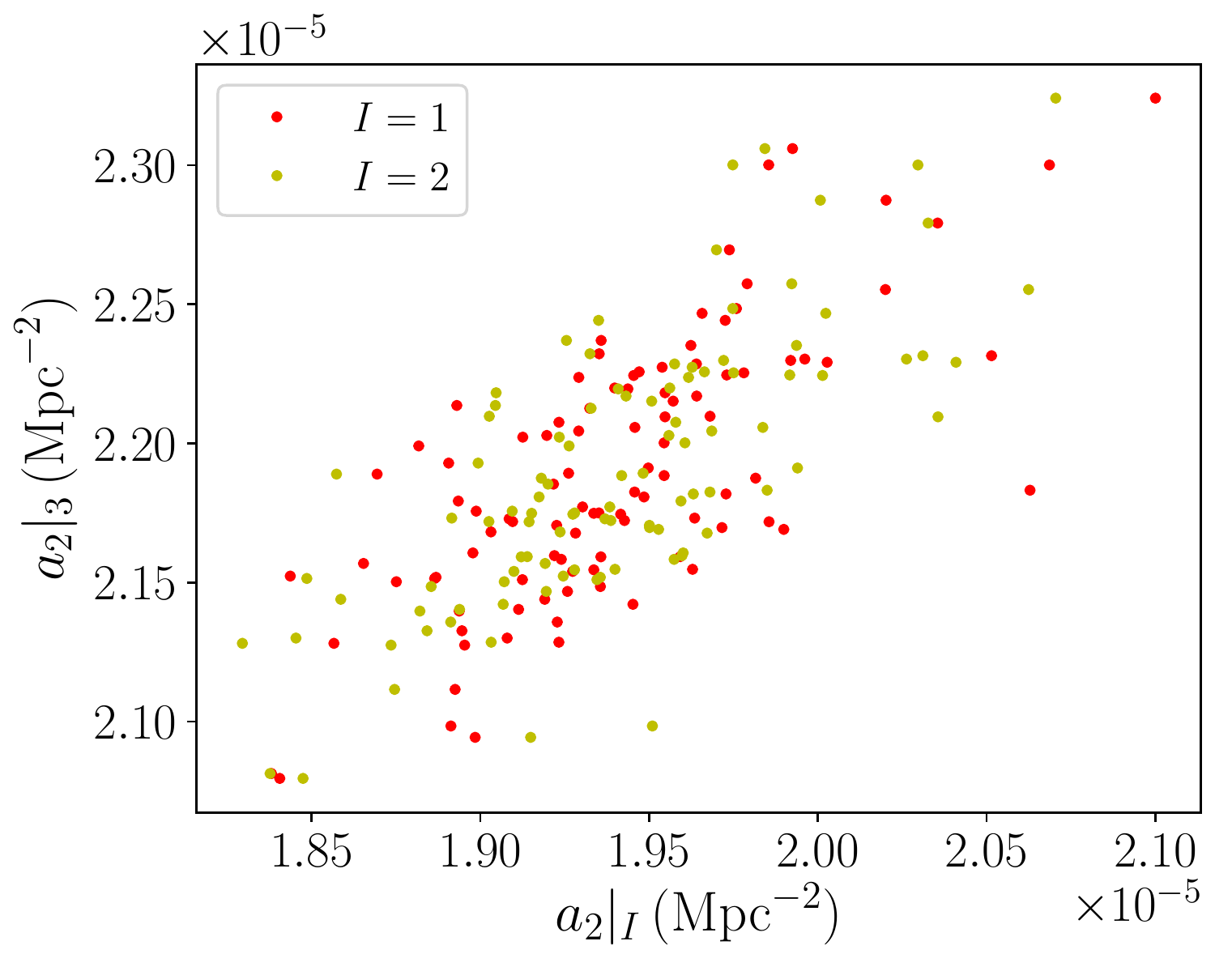}
  \caption{[Top panel] $a_{1}|_{3}$ against $a_{1}|_{I}$ measured from each $N_{\rm real} = 100$ realisation of an anisotropic Gaussian field. The red/yellow dots correspond to $I=1,2$ respectively. There is a weak positive correlation between the measurements of $a_{1}|_{3}$ against $a_{1}|_{I}$, which will be eliminated by measuring their ratio. [Bottom panel] $a_{2}|_{3}$ against $a_{2}|_{I}$ measured from each $N_{\rm real} = 100$ realisation of an anisotropic Gaussian field. The color scheme is the same as in the top panel. The correlation is significantly stronger than in the top panel. }
  \label{fig:3}
\end{figure}

One advantage of using the ratio of diagonal elements $\Theta_{1}|_{I}$, $\Theta_{2}|_{I}$ as observables is that the correlations between $a_{1}|_{I}$ and $a_{1}|_{3}$, $a_{2}|_{I}$ and $a_{2}|_{3}$ that arise as a result of being measured from the same realisation of data will be canceled. In Figure \ref{fig:3} we exhibit  $a_{1}|_{3}$ against $a_{1}|_{I}$ (top panel) and $a_{2}|_{3}$ against $a_{2}|_{I}$ (bottom panel), where each point has been extracted from a different  realisation of the anisotropic Gaussian field. The red/yellow dots are $I=1,2$  respectively. One can observe a positive correlation between these statistics, stronger for $a_{2}|_{I}$, $a_{2}|_{3}$. The linear regression slopes for the top and bottom panels are $l=0.32$ (p-value $p=0.011$) and $l=0.81$ (p-value $p \ll 0.01$) respectively. This correlated contribution to the noise will be eliminated by measuring the ratios $\Theta_{1}|_{I}$, $\Theta_{2}|_{I}$. The method of taking ratios of statistics to eliminate correlated noise follows \citep{Park:2000rc,Park:2005bu}.

It is straightforward to calculate the sensitivity of the statistics $\Theta_{1}|_{I}$, $\Theta_{2}|_{I}$ to $\beta$. We define the fractional measurement uncertainties on $\Theta_{1}|_{I}$, $\Theta_{2}|_{I}$ as $\sigma_{1I}$ and $\sigma_{2I}$ respectively, and the corresponding fractional $1-\sigma$ error on $\beta$ as $\sigma_{\beta}$. These are optimistically related by

\begin{equation}  \sigma^{2}_{\beta}  = \sigma_{1I}^{2} \left( {\partial \ln \Theta_{1}|_{I} \over \partial \ln \beta} \right)^{-2} ,  \end{equation}

\noindent For $\Theta_{1|I}$ and 

\begin{equation}   \sigma^{2}_{\beta} =  \sigma_{2 I}^{2} \left( {\partial \ln \Theta_{2}|_{I} \over \partial \ln \beta } \right)^{-2} , 
\end{equation} 

\begin{figure}
  \includegraphics[width=0.48\textwidth]{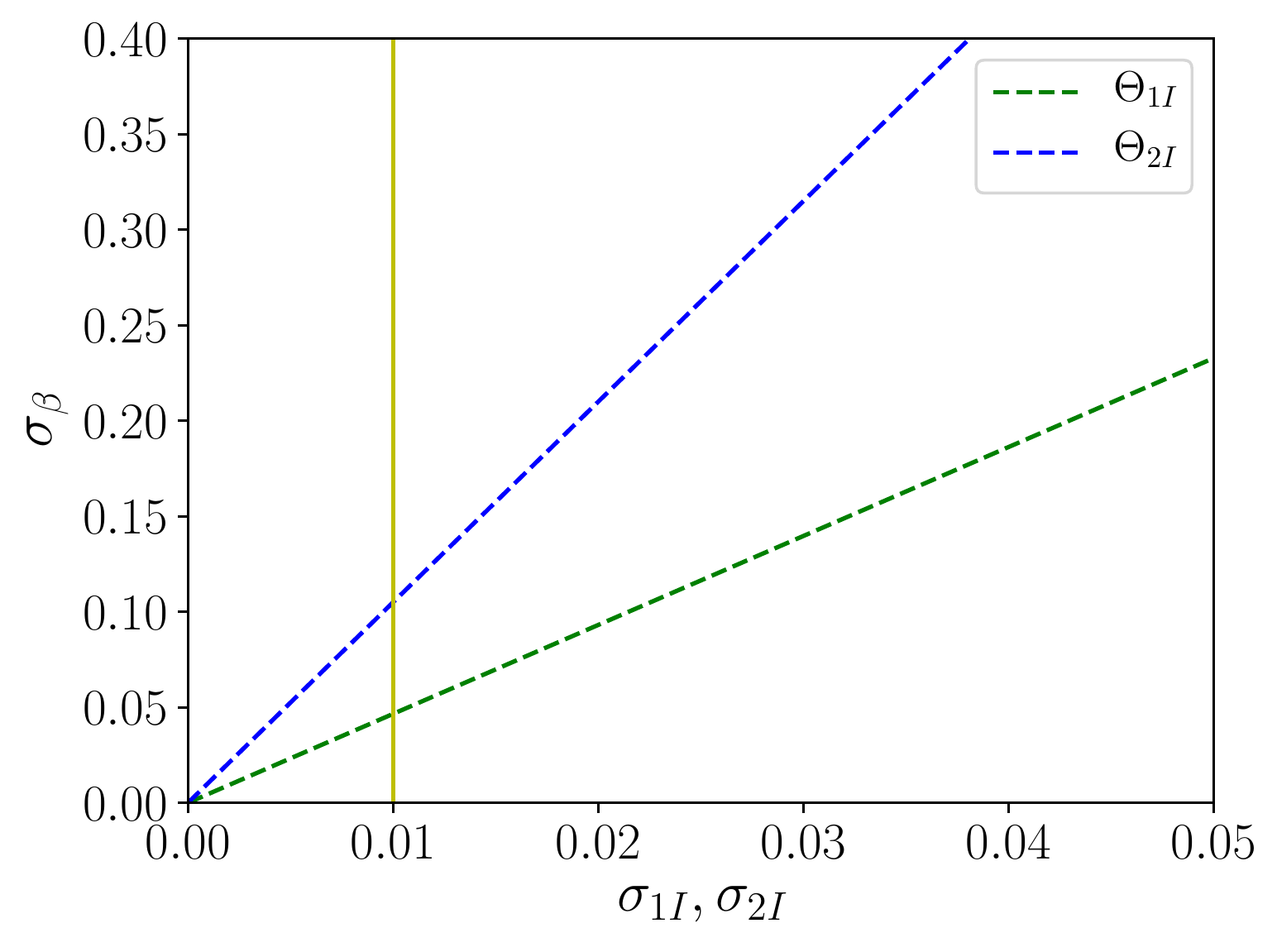}
  \caption{The fractional $1-\sigma$ uncertainty $\sigma_{\beta}$ that we could obtain on the $\beta$ parameter if we measure the statistics $\Theta_{1}|_{I}$, $\Theta_{2}|_{I}$ to accuracy $\sigma_{1I}, \sigma_{2I}$. The green/blue dashed lines correspond to $\Theta_{1}|_{I}$ and $\Theta_{2}|_{I}$ respectively. The yellow vertical line indicates a $1\%$ accurate measurement of $\Theta_{1}|_{I}$, $\Theta_{2}|_{I}$. A $1\%$ measurement of $\Theta_{1}|_{I}$ would yield a $4\%$ statistical error on $\beta$, but a $1\%$ measurement of $\Theta_{2}|_{I}$ generates a weaker, $10\%$ constraint.  }
  \label{fig:2}
\end{figure}

\noindent For $\Theta_{2|I}$. Taking a fiducial value $\beta_{\rm fid} = 0.48$, in Figure \ref{fig:2} we present $\sigma_{\beta}$ against $\sigma_{1 I}$, $\sigma_{2 I}$. This figure informs us the maximal constraint that can be obtained on $\beta$ given a measurement of $\Theta_{1}|_{I}$ (green dashed line), $\Theta_{2}|_{I}$ (blue dashed line) of accuracy $\sigma_{1 I}$, $\sigma_{2 I}$. The yellow vertical line represents a $1\%$ measurement of $\Theta_{1}|_{I}$, $\Theta_{2}|_{I}$, from which we could infer a $4\%$ or $10\%$ constraint on $\beta$ respectively. The statistic $\Theta_{1}|_{I}$, obtained from $W^{0,2}_{1}$, is more sensitive to $\beta$ and can generate tighter constraints on the redshift space distortion parameter. However, the two possess complementary information and both can be extracted from the data.

\section{Matrix Elements or Eigenvalues?} 
\label{sec:evals}

In this study we have advocated the following approach -- extract the Minkowski tensors from the data, obtain the amplitudes of the diagonal elements $a_{1}|_{I}$, $a_{1}|_{3}$, $a_{2}|_{I}$, $a_{2}|_{3}$, use these to define the ratios $\Theta_{1}|_{I}$, $\Theta_{2}|_{I}$, then compare with the theoretical expectation values ($\ref{eq:th0},\ref{eq:th1}$). For an isotropic field we should measure $\Theta_{1}|_{I} = \Theta_{2}|_{I} = 1$, but any anisotropic signal will generate a departure from unity. We now compare our approach to that proposed in \citep{Ganesan:2016jdk,K.:2018wpn}, where the Minkowski tensors were measured, the eigenvalues $\lambda_{1},\lambda_{2}$ calculated, and the ratio of eigenvalues were used to define an $\alpha = \lambda_{2}/\lambda_{1}$ parameter. We briefly explain why eigenvalues of the Minkowski Tensor matrices are problematic for extracting anisotropic signals from random fields. 

For our discussion to remain tractable, we take a $2\times 2$ matrix ${\mathcal W}$ to represent a Minkowski Tensor that can be extracted from a (possibly anisotropic) two-dimensional field. If we measure ${\mathcal W}$ from a single realisation of a two dimensional field we would obtain 

\begin{equation}
 \label{eq:f1} \mathcal W =  \left(
  \begin{array}{cc} 
    a + \epsilon \Delta_{11}  &  \epsilon \Delta_{12}  \\
   \epsilon \Delta_{12}  & b + \epsilon \Delta_{22} 
  \end{array}\right).
  \end{equation}

\noindent Here $a,b$ is the signal that we are trying to extract, and $\Delta_{11}$, $\Delta_{12}$, $\Delta_{22}$ are the statistical uncertainties of the measurement. The field is isotropic if $a=b$ and anisotropic otherwise. The magnitude of the statistical error is ${\cal O}(\epsilon)$, and $\Delta \sim {\cal O}(1)$. We assume that the $\Delta$ quantities have zero expectation value $\langle \Delta_{11}\rangle = \langle \Delta_{12}\rangle = \langle \Delta_{22}\rangle = 0$ but non-zero variances $\langle \Delta^{2} \rangle \neq 0$. The ensemble expectation value of the matrix ${\mathcal W}$ is

\begin{equation}
  \langle \mathcal W \rangle  =  \left(
  \begin{array}{cc} 
     a   &  0  \\
    0  &  b 
  \end{array}\right).
  \end{equation}

If we measure the diagonal elements of $\mathcal W$ from the data -- ${\mathcal W}_{11}, {\mathcal W}_{22}$ -- then take the ratio, we must compare the measurement to the ensemble expectation value $\langle {\mathcal W}_{11} / {\mathcal W}_{22} \rangle$, which we can write as

\begin{eqnarray} \nonumber \left\langle {{\mathcal W}_{11} \over {\mathcal W}_{22} } \right\rangle &=& \left\langle {a + \epsilon \Delta_{11} \over b + \epsilon \Delta_{22}}  \right\rangle , \\ \nonumber 
&=& {a \over b} + {\epsilon \over b} \left(\langle \Delta_{11} \rangle - {a \over b} \langle \Delta_{22}\rangle \right) + {\cal O}(\epsilon^{2})  , \\
\label{eq:c1} &=&  {a \over b} + {\cal O}(\epsilon^{2})  .
\end{eqnarray}

To linear order in the statistical uncertainty, the ratio of diagonal elements is a measure of $a/b$, the signal that we wish to extract from the data. As the field approaches the isotropic limit $a \to b$, the statistic $\langle {\mathcal W}_{11} / {\mathcal W}_{22} \rangle$ approaches unity. 

An alternative approach is to calculate the Eigenvalues $\lambda_{1}, \lambda_{2}$ of the matrix ($\ref{eq:f1}$), which for a single realisation are given by 

\begin{widetext} 

\begin{equation}   \lambda_{1,2} =  {1 \over 2} \left[ a + b + \epsilon\left(\Delta_{11} + \Delta_{22} \right) \pm \sqrt{ (a-b)^{2} + 2 \epsilon (a-b) (\Delta_{11} - \Delta_{22}) + \epsilon^{2} (\Delta_{11}^{2}+\Delta_{22}^{2} - 2 \Delta_{11}\Delta_{22} + 4 \Delta_{12}^{2}) } \right] ,
\end{equation} 

\end{widetext} 

\noindent The function $\alpha$ is defined as the ratio of eigenvalues $\alpha = \lambda_{2}/\lambda_{1}$ with $\lambda_{1} > \lambda_{2}$. Assuming a significant anisotropic signal $|b-a| \gg \epsilon$ and $b > a$, $\alpha$ can be expanded as 

\begin{equation} \alpha = {a \over b} + {\epsilon \over b}\left( \Delta_{11} - {a \over b} \Delta_{22} \right) + {\cal O}(\epsilon^{2}) , \end{equation}

\noindent and the ensemble expectation value of this statistic is therefore 

\begin{equation} \label{eq:al1} \langle \alpha \rangle =  {a \over b} + {\cal O}(\epsilon^{2}) . \end{equation}

\noindent Hence provided the field is strongly anisotropic, the statistic $\alpha$ also provides a faithful measurement of $a/b$. However, in the isotropic limit $a \to b$, $\langle \alpha \rangle$ does not smoothly approach unity. In fact for $a = b$, the expectation value of $\alpha$ is given by

\begin{equation}\label{eq:no1} \langle \alpha \rangle = 1 - {\epsilon \over a}\left\langle \sqrt{\Delta_{11}^{2} + \Delta_{22}^{2} - 2 \Delta_{11}\Delta_{22} + 4 \Delta_{12}^{2}}\right\rangle , \end{equation}

\noindent where the second term on the right hand side of ($\ref{eq:no1}$) is not generically zero for a field occupying a finite area. It approaches zero only in the limit of vanishing statistical uncertainty. 

The function $\alpha$ can differ from unity either due to an anisotropic signal $a \neq b$, or due to some non-linear function of the statistical uncertainty. As we do not have an analytic prediction for the variances $\langle \Delta^{2}_{11}\rangle$, $\langle \Delta^{2}_{12}\rangle$, $\langle \Delta^{2}_{22}\rangle$ (or any combinations thereof), and we do not know {\it a priori} the magnitude of the statistical uncertainty $\epsilon$, we cannot distinguish a true signal that we wish to measure -- $a/b$ -- from an isotropic field with a non-linear noise contribution ($\ref{eq:no1}$). 

Although the eigenvalues can conflate signal and noise, the eigen-decomposition is generally an important step in testing anisotropy with the Minkowski tensors. The principle advantage of using the eigenvalues and eigenvectors is that the eigenvalues are coordinate invariant, which is not the case for the matrix elements.  If one is studying a field or material in which the magnitude and direction of any anisotropic signal is unknown, then the first order of business should always be to calculate the eigenvalues and eigenvectors of the Minkowski tensors. Following this, one can calculate the Minkowski tensors a second time, in a coordinate system aligned with the eigenvector basis. This second step is important, as the  matrix elements provide an unbiased measure of the magnitude and significance of the anisotropic signal.

For the case of redshift space distortion, we do not need to use the eigenvectors to first generate a coordinate system aligned with the principal directions of the matrices, as we already know the direction of the anisotropic signal (parallel to the line of sight). The residual coordinate dependence in the plane perpendicular to the line of sight is irrelevant, as the matter field is invariant under rotations in this plane. For this reason, we can skip the first step and directly calculate the elements of $W^{0,2}_{1}$, $W^{0,2}_{2}$.

\section{Discussion}
\label{sec:5}

The main results of this work are equations ($\ref{eq:co1}$-$\ref{eq:co4}$), which are the ensemble expectation values of $W^{0,2}_{1}$, $W^{0,2}_{2}$ extracted from the matter density field in redshift space, in a coordinate system aligned with the line of sight. After reviewing the calculation of these statistics for an isotropic field, we considered the generalisation in which a global anisotropic signal is generated by the velocity component along the line of sight. Under the assumption that all fields are Gaussian, we calculated the effect of the velocity field on the Minkowski tensors. We find the statistics remain diagonal, however the elements of the matrix are no longer equal. The ratio of the Minkowski Tensor diagonal elements parallel and perpendicular to the line of sight -- $\Theta_{1}|_{I}$, $\Theta_{2}|_{I}$ -- are functions only of the redshift space distortion parameter $\beta$. It follows that by measuring these quantities from galaxy data (for example), one can obtain a pristine constraint on the growth rate of density perturbations. 

We numerically confirmed our analytic results, and then estimated the sensitivity of $\Theta_{1}|_{I}$, $\Theta_{2}|_{I}$ to the parameter $\beta$, finding that a $1\%$ accurate reconstruction of these statistics from a data set can yield a $4\%$ or $10\%$ constraint on $\beta$ respectively. Our method will be competitive with current state of the art measurements involving the velocity field $\beta(z \simeq 0.025) = 0.49^{+0.08}_{-0.05}$ \citep{Park:2005bu} and void-galaxy cross-correlation $\beta(z=0.54) = 0.457^{+0.056}_{-0.054}$ \citep{Hamaus:2017dwj}. 

It is important to stress that the Minkowski tensors are not coordinate invariant. Specifically, the Minkowski tensors used in this work are translation, but {\it not}  rotation, invariant.  As such, the analytic results obtained here are only applicable in the distant observer limit, and in a coordinate system aligned with the line of sight. For a field in which an anisotropic signal may be present but its direction unknown, we advocate measuring the Minkowski tensors in an arbitrary coordinate system, constructing their eigenvectors, and then re-calculating the Minkowski tensors in this basis. For the case of redshift space distortion, the eigen-decomposition is unnecessary as the direction of the anisotropic signal is already known. 

The calculation undertaken in this work applies to linear redshift space distortions only. The matter density field in the low redshift Universe is non-Gaussian as a result of gravitational collapse. For a non-Gaussian but isotropic field, the Minkowski tensors will remain proportional to the identity matrix. However, when non-Gaussian velocity corrections are accounted for, one can expect modifications to the Gaussian predictions ($\ref{eq:m1}$-$\ref{eq:m3}$), ($\ref{eq:m4}$-$\ref{eq:m6}$). A detailed study of the the non-Gaussian regime will be considered elsewhere.

\section*{Acknowledgement}

We thank the Korea Institute for Advanced Study for
providing computing resources (KIAS Center for Advanced
Computation Linux Cluster System).

The work of P. Chingangbam is supported by the Science and Engineering Research Board of the Department of Science and Technology, India, under the \texttt{MATRICS} scheme, bearing project reference no \texttt{MTR/2018/000896}.

\section*{Appendix A -- Mean Curvature Contributions to Ensemble Expectation}

In the appendix we show that all terms proportional to $y_{ii}$ in $G_{2}$ integrate to zero, and hence do not contribute to $\langle W^{0,2}_{2} \rangle$. In the isotropic case, the terms that we consider are defined as the matrix ${\mathcal T}_{lm}$ as 

\begin{widetext} 

\begin{equation} \label{eq:eq1} \langle {\mathcal T}_{lm} \rangle =  \int_{-\infty}^{\infty} dx_{1} \int_{-\infty}^{\infty} dx_{2} \int_{-\infty}^{\infty}dx_{3} e^{-(x_{1}^{2} + x_{2}^{2} + x_{3}^{2})/2} \int d{\bf y} e^{-{\bf y}^{\rm T} \Sigma^{-1} {\bf y}/2} \left[ { x_{1}^{2} (y_{22} + y_{33}) + x_{2}^{2}(y_{11}+y_{33}) + x_{3}^{2} (y_{11}+y_{22}) \over X^{4}} \right] {\mathcal M}_{lm} ,
\end{equation}

\end{widetext}

\noindent where we have defined the three-dimensional vector ${\bf y} = (y_{11},y_{22},y_{33})$ with associated covariance

\begin{equation}
  \Sigma  =  {1 \over 3} \left(
  \begin{array}{ccc} 
    3 & 1  &  1  \\
        1 & 3  &  1  \\
    1 & 1  &  3    \end{array}\right),
  \end{equation}
  
  \noindent and have set any constant multiplicative factors to unity. ${\mathcal T}_{lm}$ is the contribution to $\langle W^{0,2}_{2} \rangle $ arising from the $y_{ii}$ terms in $G_{2}$. 
  
  The off-diagonal elements of $\langle {\mathcal T}_{lm} \rangle$ are zero as the matrix ${\mathcal M}_{lm}$ is an odd function of $x_{1}$, $x_{2}$, $x_{3}$ for $l \neq m$. To verify that the diagonal elements $\langle {\mathcal T}_{ll}\rangle $ are also zero, we simply show that there exists an invertible linear transformation ${\bf z} = L {\bf y}$ which renders the covariance matrix $\Sigma$ diagonal. If so, the exponential transforms as $e^{{-\bf y}\Sigma^{-1}{\bf y}/2} \to e^{-(z_{11}^{2}+z_{22}^{2} + z_{33}^{2})/2}$ and the terms in the integrand linearly related to $y_{11}$, $y_{22}$, $y_{33}$ will be transformed to contributions linear in $z_{11}, z_{22}, z_{33}$.  
  
  One such transformation is given by \citep{1986ApJ...304...15B}

 \begin{equation}
  L  =  \left(
  \begin{array}{ccc} 
    1/\sqrt{5} & 1/\sqrt{5}  &  1/\sqrt{5}  \\
        \sqrt{3}/2 & 0 &  -\sqrt{3}/2  \\
    1/2 & -1  &  1/2    \end{array}\right),
  \end{equation}
  
 \noindent which transforms $\Sigma$ to the identity matrix. The terms linear in $y_{ii}$ in ($\ref{eq:eq1}$) are transformed to terms linear in $z_{ii}$ according to ${\bf y} = L^{-1} {\bf z}$, and the integrals are zero by virtue of the integrand being an odd function of $z_{11}, z_{22}, z_{33}$.
 
 Similarly, for an anisotropic field we must verify that all terms proportional to $y_{II}$, $y_{33}$ in $G_{2}$ integrate to zero. Defining ${\mathcal G}_{2}$ as the terms in $G_{2}$ that we wish to confirm do not contribute to the expectation value of $\langle W^{0,2}_{2} \rangle$ as -- 
 
  \begin{widetext} 

\begin{equation}
     {\mathcal G}_2 =  -{\sqrt{3}  \over 2\sqrt{5}\sigma_{1 \perp}\left(x_1^2+x_2^2+\lambda^{2} x_3^2\right)^{3/2}}\left[ \sigma_{2\perp}\sqrt{1-\gamma_{\perp}^{2}} \left[ (x_{2}^{2} + \lambda^{2}x_{3}^{2})y_{11} + (x_{1}^{2} + \lambda^{2}x_{3}^{2})y_{22} \right] + \sigma_{2\parallel}\sqrt{1 - \gamma_{\parallel}^{2}}(x_{1}^{2} + x_{2}^{2})y_{33} \right] .  
\end{equation}

\end{widetext}

\noindent This function will contribute the following to the ensemble expectation $\langle W^{0,2}_{2} \rangle$ --

\begin{widetext} 

\begin{eqnarray} \nonumber  \langle \tilde{{\mathcal T}}_{lm} \rangle  &=& \int_{-\infty}^{\infty} dx_{1} \int_{-\infty}^{\infty} dx_{2} \int_{-\infty}^{\infty}dx_{3} e^{-(x_{1}^{2} + x_{2}^{2} + x_{3}^{2})/2} \int d{\bf y} e^{-{\bf y}^{\rm T} \tilde{\Sigma}^{-1} {\bf y}/2} {\tilde{{\mathcal M}} \over \tilde{X}^{4}} \times  \\
\label{eq:eq2} & & \qquad\left[ \sigma_{2\perp}\sqrt{1-\gamma_{\perp}^{2}} \left[ (x_{2}^{2} + \lambda^{2}x_{3}^{2})y_{11} + (x_{1}^{2} + \lambda^{2}x_{3}^{2})y_{22} \right] + \sigma_{2\parallel}\sqrt{1 - \gamma_{\parallel}^{2}}(x_{1}^{2} + x_{2}^{2})y_{33} \right]   ,
\end{eqnarray}

\end{widetext}

\noindent where the covariance $\tilde{\Sigma}$ in redshift space is given by 

\begin{equation}
  \tilde{\Sigma}   =  {1 \over 3} \left(
  \begin{array}{ccc} 
    3 & 1  &  3\kappa^{2}  \\
        1 & 3  &  3\kappa^{2}  \\
    3\kappa^{2} & 3\kappa^{2}  &  3    \end{array}\right),
  \end{equation}

\noindent and we have defined $\kappa^{2} = \sigma_{2\times}^{2}/(3\sigma_{2\perp}\sigma_{2\parallel})$. 

The off-diagonal elements of $\langle \tilde{\mathcal T}_{lm} \rangle$ are zero as in the isotropic case, and finding an invertible linear transformation ${\bf z} = \tilde{L} {\bf y}$ that renders $\tilde{\Sigma}$ diagonal will show that the diagonal elements $\langle \tilde{{\mathcal T}}_{ll} \rangle$ are also zero. A suitable linear transformation $\tilde{L}$ is given by 
  
  \begin{widetext} 
  
  \begin{equation}
  \tilde{L}  =  \left(
  \begin{array}{ccc} 
    \sqrt{3/8} & \sqrt{3/8}  &  0  \\
        \sqrt{3}/2 & -\sqrt{3}/2 &  0  \\
    -3\kappa^{2}/(4\sqrt{1-3\kappa^{4}/2}) & -3\kappa^{2}/(4\sqrt{1-3\kappa^{4}/2})  &  1/\sqrt{1-3\kappa^{4}/2}    \end{array}\right).
  \end{equation}

 \end{widetext} 
 
 \noindent This transformation renders the integrand in ($\ref{eq:eq2}$) an odd function of $z_{11}$, $z_{22}$, $z_{33}$, confirming that $\langle \tilde{{\mathcal T}}_{lm} \rangle  = 0$.

\bibliography{biblio}{}

\end{document}